\documentclass{emulateapj}

\slugcomment{ver.Jan.6; draft 1 pages}
\slugcomment{ver.Mar.30; draft 7 pages}
\slugcomment{ver.Jun.25; draft 13 pages}
\slugcomment{ver.Jul.3; draft v1.1}
\slugcomment{ver.Jul.26; draft v2}
\slugcomment{ver.Jul.31; draft v3, submitted}
\slugcomment{ver.4: 2012.08.24; revisions after the referee's comments}
\slugcomment{ver.5: 2012.09.12; revisions after the referee's comments}
\slugcomment{ver.6: 2012.10.15; revisions after the 2nd referee's comments}
\slugcomment{Received 2012 July 31; accepted 2012 October 15}

\shorttitle{A LAE with EW$_0\sim900$\AA}
\shortauthors{Kashikawa et al.}

\begin{document}


\title{A Ly$\alpha$ Emitter with an Extremely Large Rest-frame Equivalent Width of $\sim900$\AA~at $z=6.5$: A Candidate of Population III-dominated Galaxy?\altaffilmark{1}}


\author{
Nobunari Kashikawa\altaffilmark{2,3}, 
Tohru Nagao\altaffilmark{4,5}, 
Jun Toshikawa\altaffilmark{3}, 
Yoshifumi Ishizaki\altaffilmark{3}, 
Eiichi Egami\altaffilmark{6}, 
Masao Hayashi\altaffilmark{2}, 
Chun Ly\altaffilmark{7,12}, 
Matthew A. Malkan\altaffilmark{8}, 
Yuichi Matsuda\altaffilmark{9}, 
Kazuhiro Shimasaku\altaffilmark{10}, 
Masanori Iye\altaffilmark{2,3}, 
Kazuaki Ota\altaffilmark{4}, 
Takatoshi Shibuya\altaffilmark{3}, 
Linhua Jiang\altaffilmark{6}, 
Yoshiaki Taniguchi\altaffilmark{11}, 
and
Yasuhiro Shioya\altaffilmark{11}
}

%
%
%
%Kentaro Motohara\altaffilmark{8}, 
%Masami Ouchi\altaffilmark{10}, 
%Hisanori Furusawa\altaffilmark{14}, 
%Takashi Hattori\altaffilmark{12}, 
%Sadanori Okamura\altaffilmark{5,6}, 
% M Hayashi
%Kouji Ohta\altaffilmark{9}, 
%Mamoru Doi\altaffilmark{8}, 
%Takashi Murayama\altaffilmark{11}, 
%, and
%Masayuki Umemura\altaffilmark{15}
%Makiko Yoshida\altaffilmark{4}, 
%Masahiro Nagashima\altaffilmark{6}, 
%Hideki Yahagi\altaffilmark{4}, 
%Masayuki Akiyama\altaffilmark{12}, 
%Hiroyasu Ando\altaffilmark{2}, 
%Kentaro Aoki\altaffilmark{12}, 
%%Shinobu S. Fujita\altaffilmark{13}, 
%Tomoki Hayashino\altaffilmark{13}, 
%Fumihide Iwamuro\altaffilmark{14}, 
%Hiroshi Karoji\altaffilmark{14}, 
%Naoto Kobayashi\altaffilmark{10}, 
%Keiichi Kodaira\altaffilmark{15}, 
%Tadayuki Kodama\altaffilmark{2}, 
%Yutaka Komiyama\altaffilmark{2}, 
%Satoshi Miyazaki\altaffilmark{12}, 
%Tomoki Morokuma\altaffilmark{10}, 
%Toru Yamada\altaffilmark{2}, and 
%Naoki Yasuda\altaffilmark{19}

\email{n.kashikawa@nao.ac.jp}


%\altaffiltext{1}{The data presented herein were partly obtained at the W. M. Keck Observatory, which is operated as a scientific partnership among the California Institute of Technology, the University of California, and the National Aeronautics and Space Administration. The Observatory was made possible by the generous financial support of the W. M. Keck Foundation.}
\altaffiltext{1}{Based on data collected at the Subaru Telescope, which is operated by the National Astronomical Observatory of Japan, and the W. M. Keck Observatory, which is operated as a scientific partnership among the California Institute of Technology, the University of California, and the National Aeronautics and Space Administration. The Observatory was made possible by the generous financial support of the W. M. Keck Foundation.}
\altaffiltext{2}{Optical and Infrared Astronomy Division, National Astronomical Observatory, Mitaka, Tokyo 181-8588, Japan.}
\altaffiltext{3}{Department of Astronomy, School of Science, Graduate University for Advanced Studies, Mitaka, Tokyo 181-8588, Japan.}
\altaffiltext{4}{Department of Astronomy, Graduate School of Science, Kyoto University, Kyoto 606-8502, Japan.}
\altaffiltext{5}{The Hakubi Center for Advanced Research, Kyoto University, Kyoto 606-8302, Japan.}
\altaffiltext{6}{Steward Observatory, University of Arizona, 933 North Cherry Avenue, Tucson, Arizona 85721, USA}
\altaffiltext{7}{Space Telescope Science Institute, 3700 San Martin Drive, Baltimore, MD 21218.}
\altaffiltext{8}{Department of Physics and Astronomy, University of California, Los Angeles, CA 90095-1547.}
\altaffiltext{9}{Radio Astronomy Division, National Astronomical Observatory, Mitaka, Tokyo 181-8588, Japan.}
\altaffiltext{10}{Department of Astronomy, University of Tokyo, Hongo, Tokyo 113-0033, Japan.}

\altaffiltext{11}{Research Center for Space and Cosmic Evolution, Ehime University, Bunkyo-cho, Matsuyama 790-8577, Japan.}
\altaffiltext{12}{Giacconi fellow.}

\begin{abstract}
We have identified a very interesting Ly$\alpha$ emitter, whose Ly$\alpha$ emission line has an extremely large observed equivalent width of EW$_0=436^{+422}_{-149}$\AA~, which corresponds to an extraordinarily large intrinsic rest-frame equivalent width of EW$_0^{int}=872^{+844}_{-298}$\AA~ after the average intergalactic absorption correction.
The object was spectroscopically confirmed to be a real Ly$\alpha$ emitter by its apparent asymmetric Ly$\alpha$ line profile detected at $z=6.538$.
%The EW$_0$ of most of Pop III candidates identified in previous works were only given by a lower limit because their continuum emission were not detected, while 
The continuum emission of the object was definitely detected in our deep $z'$-band image; thus, its EW$_0$ was reliably determined.
Follow-up deep near-infrared spectroscopy revealed emission lines of neither He {\sc ii} $\lambda1640$ as an apparent signature of Population III, nor C {\sc iv} $\lambda1549$ as a proof of active nucleus.
No detection of short-lived He {\sc ii} $\lambda1640$ line is not necessarily inconsistent with the interpretation that the underlying stellar population of the object is dominated by Population III.
We found that the observed extremely large EW$_0$ of the Ly$\alpha$ emission and the upper limit on the EW$_0$ of the He {\sc ii} $\lambda1640$ emission can be explained by population synthesis models favoring a very young age less than $2-4$Myr and massive metal-poor ($Z<10^{-5}$) or even metal-free stars.
The observed large EW$_0$ of Ly$\alpha$ is hardly explained by Population I/II synthesis models with $Z\geq10^{-3}$.
However, we cannot conclusively rule out the possibility that this object is composed of a normal stellar population with a clumpy dust distribution, which could enhance the Ly$\alpha$ EW$_0$, though its significance is still unclear.

%Such an enormous EW$_0^{int}\sim900$\AA~in the Ly$\alpha$ emission cannot be attained by the Population II synthetic cluster, and only achieved in the case of Pop III with ages as young as $<2$Myr.
%We propose deep MOIRCS spectroscopy to detect its He {\sc ii} emission line at $12373.8$\AA, which would definitely represent the first direct detection of Pop III-dominated object.
%This exciting discovery will exactly identify the critical epoch when the universe begins to be enriched by metals, and must have a strong impact on theoretical predictions about IMF, mass loss strength and age of Pop III.
%These will be discussed along with independent estimates of stellar mass, age, and dust extinction by our upcoming deep HST-NICMOS+Spitzer joint observation targeting our $z=6.5$ LAE sample, which includes SDF-popIII-1.

\end{abstract}




%% \citep　()あり
%% and 
%% \citet ()なし
%% and
%% \citealp ()なし年もなし


\section{Introduction}

Big-bang nucleosynthesis could not produce elements heavier than lithium at the beginning, while today we live in a universe with $90$ kinds of constituent natural elements.
It is thought that the universe was first metal enriched by the first generation of stars, Population III (Pop III) stars.
Moreover, Pop III may have played a key role in cosmic reionization, and observational constraints on their properties will significantly help the theoretical model to predict their characteristics as well as the quantity of ionizing photons produced by these first stars.

The observational characteristics to identify the first stars have been discussed by \citealp{tum01}, \citealp{bro01}, \citealp{oh01}, \citealp{sch02}, \citealp{sch03}, and \citealp{tum03}.
The first stars were born in an extreme condition with a metal-free, and presumably top-heavy initial mass function (IMF), though it held much in doubt by recent simulations (e.g., \citealp{kru09}, \citealp{tur09}, \citealp{sta12}, \citealp{hos11})
Their metal ejection to the intergalactic medium (IGM) may have caused a significant change in subsequent star formation.
In the pristine gas cloud with low metallicities $Z \leq Z_{crit}=10^{-5\pm1}Z_\odot$, molecules such as H$_2$ or HD dominated the cooling.
Since zero-metal Pop III stars cannot burn in the CNO cycle like normal massive stars, their energy production has to rely initially on inefficient proton-proton burning. 
Therefore these stars have higher core temperatures of $\sim10^8$K, which makes them hotter and smaller than their metal-enriched counterparts.
Given the exceptionally high effective temperatures of Pop III stars in the zero-age main sequence, they emit a larger fraction of the luminosity in the Lyman continuum and have a much harder ionizing spectrum than stars with higher metallicity.
The main characteristics of the predicted spectral energy distribution (SED) are the presence of a strong Ly$\alpha$ emission line due to the strong ionizing flux and a He$^+$ recombination line (especially He {\sc ii} $\lambda1640$) due to spectral hardness.

Despite great progress in the theoretical prediction of the unique physical properties of Pop III, direct observational detection of Pop III stars has yet to be achieved.
Based on a model prediction of Pop III abundances \citep{sca03}, it is suggested that part of Pop III may have already been detected in ongoing surveys for high-$z$ Lyman-$\alpha$ emitters (LAEs).
There is indirect observational evidence for the existence of Pop III.
A relatively high median value ($\sim240$\AA) was derived from the rest-frame equivalent width (EW$_0$)\footnote{We refer to the rest-frame equivalent width, which is a factor of ($1+z$) lower than the observed equivalent width.} distribution of LAEs at z$=4.5$ \citep{mal02}. 
However, the observational estimate of EW$_0$ at high-z is quite uncertain, because the continuum emission is faint or non-detected.
A weak He {\sc ii} emission signature was detected in the composite spectrum of Lyman-break galaxies at $z\sim3$ \citep{jim06, sha03} and $z\sim4$ \citep{jon12}, although in these composite spectra, the expected nebular He {\sc ii} feature is indistinguishable from the stellar wind origin associated with evolved massive stars.
%while no evidences of He {\sc ii} emission was found in the composite spectrum of LAEs at z=xxx (\citealp{daw04}, \citealp{ouc08})
\citet{ino11} found some LAEs at $z=3.1$ with extremely strong Lyman continuum flux, which can be explained by a very young and massive metal-poor or metal-free stellar population, if the mass fraction of such a population is $\sim1-10\%$ of the total stellar mass of the LAEs.
More straightforward but challenging attempts have failed to obtain direct evidence of Pop III, identifying simultaneously both Ly$\alpha$ and He {\sc ii} emissions in an individual spectrum \citep{nag05, cai11} or using a couple of narrow-band filters whose wavelengths are matched to the redshifted Ly$\alpha$ and He {\sc ii} emissions \citep{nag08}.
%This proposal is complementary to another unique approach of Nagao et al. who are going to search for Pop III objects at $z\sim4.5$ among Ly$\alpha$-He {\sc ii} dual-emitter candidates using a couple of narrow-band images.

We have found a very plausible Pop III candidate, which has an extraordinary large intrinsic EW$_0^{int}$ \footnote{We refer to the IGM-attenuation corrected EW$_0$.}
of $\sim900$\AA~in the Ly$\alpha$ emission line at $z\sim6.5$.
Such an enormous EW$_0^{int}\sim900$\AA~in the Ly$\alpha$ emission cannot be attained by the Population II synthetic cluster, and it is expected to be achieved by Pop III (with a Salpeter IMF up to $100M_\odot$) stars with young ages (\citealp{sch03}, \citealp{tum03}).
Based on these models, the expected EW$_0$ of the He {\sc ii} emission should be as large as $\sim100$\AA~for the object.
In this study, we take a deep NIR spectroscopy for the candidate to verify its redshifted He {\sc ii} $\lambda1640$ emission line at $12362$\AA~as firm evidence of a Pop III-dominated object.
The detection of He {\sc ii} emission on the spectrum of this object will definitely provide the direct evidence of a Pop III-dominated object, as well as information about the critical epoch when the universe began to be enriched by metals.
He {\sc ii} detection with a secure Ly$\alpha$ EW$_0$ estimate will have a strong impact on theoretical predictions about IMF, metallicity and the age of the metal-poor population.
%mass loss strength, and age of Pop III.
The feedback effects from Pop III would deeply affect subsequent star-formation, initial galaxy-formation, and IGM evolution (e.g., \citealp{cia07}).
%These will be discussed along with independent estimates of stellar mass, age, and dust extinction by the on-going deep HST-NICMOS+Spitzer joint observation (PI: Egami, E.) targeting our $z=6.5$ LAEs, including SDF-popIII-1.

%Even if we could not detect He {\sc ii} emission, the proposed observation will give an upper limit of He {\sc ii} flux, which will also constrain the Pop III-dominated fraction (Nagao et al. 2005), as well as metallicity as a PopII/I-dominated object.
%null resultの時はHeII fluxのupper limitが求められ、それからSchererのモデるのPOPIIIの貢献度、popIIの場合のIMF、 mass lossの大小、ageなどに制限を与えられる。それはスケジュールされているNICMOS/Spitzerの観測から得られるLAEの年齢、stellar massと比較できる。
An alternative interpretation of a large Ly$\alpha$ EW$_0$ is that the target has a large AGN contribution.
We could clearly distinguish the AGN interpretation from a Pop III origin by identifying the C {\sc iv} $\lambda1549$ emission line as an apparent AGN signature at $11676$\AA~along with He {\sc ii}.
The AGN fraction of the high-$z$ LAE population at $z=6.5$ is also an important question (\citealp{dij06}, \citealp{wan04}).
The He {\sc ii} line can also originate in the hot, dense stellar winds of Wolf-Rayet stars, although the expected He {\sc ii} EW$_0$ would be small (e.g., $2.7$\AA~in the case of \citealp{erb10}) compared with the predictions of Pop III stars.
In the case, detection of a broad He {\sc ii} line ($v_{\rm FWHM}\sim1500$km s$^{-1}$) and a simultaneous stronger C {\sc iv} $\lambda1549$ emission line with a P-Cygni profile would be expected \citep{lei95}.
%In addition, the object will be identified as the highest-$z$ AGN.
Otherwise, the large Ly$\alpha$ EW$_0$ could be the result of scattering in a clumpy, dusty medium \citep{neu91, han06}.
%, though this is a very extreme explanation.
%This interpretation will also be compared by the degree of dust extinction evaluated from NICMOS/Spitzer observation.

%Our approach here is more conservative as we are searching for candidates with much larger EW$_0$ at high-$z$.

%If it proved successful to directly detect the He {\sc ii} emission from the object, we will move on the sunsequent systematic search for Pop III by follow-up NIR spectroscopy for other secoundary plausible targets having a high EW$_0$ in the Ly$\alpha$ emission 

This paper is organized as follows:
In \S~2, we describe the photometric and spectroscopic identification of the candidate, and its Ly$\alpha$ EW$_0$ estimate.
In \S~3, we describe our deep NIR spectroscopic observation aimed at detecting its He {\sc ii} $\lambda1640$ or C {\sc iv} $\lambda1549$ emissions.
In \S~4, we present the measurements of flux upper limits on these lines.
Several possible interpretations of the observed large Ly$\alpha$ EW$_0$ are discussed in \S~5.
%We present the composite spectrum of our spectroscopically confirmed z6p5LAE sample in \S~5.
A summary of the paper is provided in \S~6, with some discussion of the implications of our results.

Throughout the paper, we assume cosmology parameters: $\Omega_{\rm m}=0.3$, $\Omega_\Lambda=0.7$, and $H_0=70$ $h_{70}$ km s$^{-1}$ Mpc$^{-1}$. 
These parameters are consistent with recent CMB constraints \citep{kom09}.
Magnitudes are given in the AB system.

\section{The Candidate and its Ly$\alpha$ Equivalent width}

\subsection{Optical Spectroscopy}

\begin{figure}
\epsscale{1.20}
\plotone{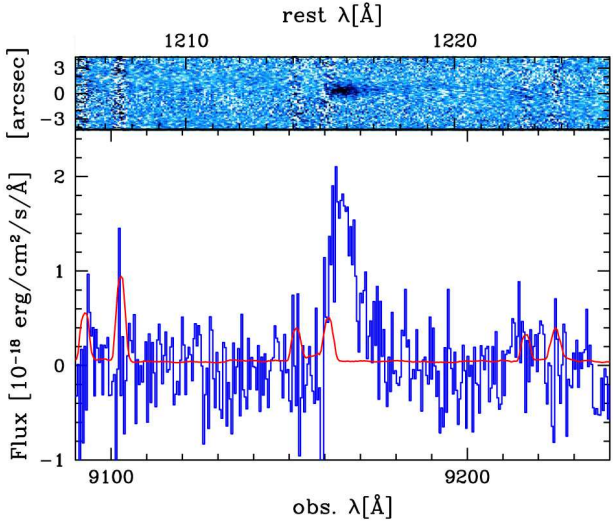}
%\vspace*{-0.8cm}
\caption{Two-dimensional (upper) and one-dimensional (lower) optical spectrum of SDF-LEW-1. 
Red line represents sky lines.
\label{fig_optspec}}
\end{figure}

The object with a large Ly$\alpha$ EW, SDF-LEW-1, was discovered in the course of our systematic large spectroscopic survey of LAEs at $z=6.5$ \citep{kas06, kas11} in the Subaru Deep Field (SDF).
We have identified a total of $43$ spectroscopically confirmed LAEs at $z=6.5$ \citep{kas11}.
The sample was based on the flux excess objects in the narrowband $NB921$ ($\lambda_c=9196$ \AA, FWHM=$132$ \AA) image, compared with the deep broadband images of the SDF.
SDF-LEW-1 was first spectroscopically confirmed to be a real LAE by Subaru/FOCAS observation on May 20-21 2007, though the quality of spectra was not good due to poor conditions.
% 14.4ksec, 0."7
The object was re-observed by Keck/DEIMOS on April 27, 2009.
The integration time was $10.8$ksec with a seeing size of $0.\arcsec7$.
The object shows an apparent asymmetric Ly$\alpha$ line profile detected at $z=6.538$\footnote{The spectroscopic properties of the object, SDF J132458.0+272349, are slightly different from those listed in Table 2 of \citet{kas11}, which is based on the previous Subaru/FOCAS low-quality spectrum. We use measurements based on the DEIMOS observation throughout this paper.}(Figure \ref{fig_optspec}).
To quantitatively estimate the line asymmetry, we introduce the weighted skewness parameter, $S_W$ \citep{kas06}, which is defined as the third moment of flux distribution multiplied by the line width.
The $S_W$ of SDF-LEW-1 is estimated to be $S_W=9.33\pm0.27$, which is larger than the empirical critical value of $S_W=3$\AA~to distinguish Ly$\alpha$ emission from other emission lines at $z>5.7$.
The clear asymmetry of the emission line ensures that it is certainly a Ly$\alpha$ emission, distinct from foreground nebular emissions.
The high resolving power of DEIMOS can distinguish a single Ly$\alpha$ emission from $[$O {\sc ii}$]$ doublets (at a rest-frame separation of $2.78$\AA), and the absence of any other emission line features both in the optical, and the NIR spectrum, as will be shown later in Section 4, rules out the possibility of its being some other emission line, such as H$\alpha$, H$\beta$, and $[$O {\sc iii}$]$.
It has a total Ly$\alpha$ luminosity of ($1.07\pm0.30)\times 10^{43}$ erg s$^{-1}$.
This corresponds to a Ly$\alpha$-based star formation rate (SFR) of $9.7$ M$_\odot$ yr$^{-1}$ using SFR(Ly$\alpha$)$=9.1\times10^{-43}L$(Ly$\alpha$) M$_\odot$ yr$^{-1}$ \citep{ken98}, though it is more or less affected by uncertainties of dust extinction, IMF, and Ly$\alpha$ escape fraction.

\subsection{EW$_0$ measurement}

\begin{figure}
\epsscale{1.1}
\plotone{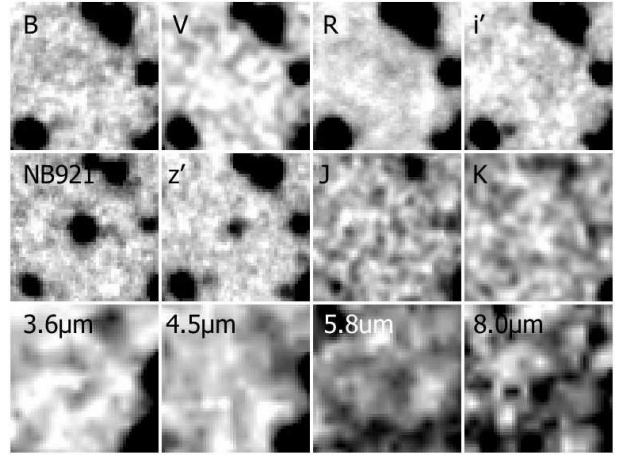}
\caption{Thumbnail images of SDF-LEW-1. The $B$, $V$, $R$, $i'$, $NB921$, $z'$, $J$, $K$, $3.6\mu$m, $4.5\mu$m, $5.8\mu$m, and $8.0\mu$m-band images are shown from upper-left to lower-right.
Each image is 10\arcsec on a side.
North is up, and east is to the left.
A constant gray level corresponds to constant $f_\nu$ surface brightness.
\label{fig_stamp}}
\end{figure}

EW$_0$ was calculated as in \citet{kas11}, using narrow- ($NB921$) and broad-band ($z'$) photometry taking into account the non top-hat shape bandpass, which potentially has substantial consequences in deriving line and continuum fluxes \citep{gro07}.
We assumed a simple model spectrum with constant UV continuum and Ly$\alpha$ emission line, including complete IGM absorption at the blue ward of Ly$\alpha$.
The Ly$\alpha$ emission flux, $f^{phot}$ (Ly$\alpha$), and the UV continuum flux density at the Ly$\alpha$ wavelength $f^{phot}_{cont}$($0.92\micron$), can be separately estimated by fitting the model spectrum to the observed fluxes in the narrow-band and the $z'$-band.
Table \ref{tab_phot} summarizes the photometry, and Figure \ref{fig_stamp} shows thumbnail images of SDF-LEW-1.
SDF-LEW-1 was only detected in $NB921$- and $z'$-band images, and it was below the $3\sigma$ limiting magnitude at other wavelengths.
The errors in these photometrical fluxes are estimated in the same way as \citet{kas11}, in which Gaussian random photometric errors in both the narrow- and broad-band were assigned to the measured magnitudes. 
The rms fluctuations of the fluxes are estimated from a Monte Carlo simulation.
Although we obtained multi-wavelength photometric data for the object, it is difficult to put a meaningful constraint on its stellar population, even for its stellar mass, by SED fitting because the object is detected only in $NB921$- and $z'$-band images.
We confirmed that the photometrically determined Ly$\alpha$ emission flux, $f^{phot}$(Ly$\alpha$)$=2.34^{+0.06}_{-0.06}\times10^{-17}$ergs s$^{-1}$ cm$^{-2}$, is fairly consistent with that measured in the optical spectrum, $f^{spec}_{opt}$(Ly$\alpha$)$=(2.19\pm0.62)\times10^{-17}$ergs s$^{-1}$ cm$^{-2}$, where errors only reflect spectroscopic rms uncertainties.
Simultaneously, the UV continuum emission flux density at Ly$\alpha$ wavelength is estimated to be $f^{phot}_{cont}$($0.92\micron$)$=(7.12\pm3.41)\times10^{-21}$ergs s$^{-1}$ cm$^{-2}$ \AA$^{-1}$, though its flux is as small as $2\sigma$ level.
This implies that most of the $z'$-band flux would be contributed by the Ly$\alpha$ flux.
Therefore, the EW$_0$ of SDF-LEW-1 is estimated to be $436^{+422}_{-149}$\AA.
We here assumed the continuum SED to be $f_\lambda \propto \lambda^\beta$ and $\beta=-2.0$.
\citet{fin11c} found that the UV continuum slope $\beta$ of Lyman Break Galaxies (LBGs) evolves significantly from $\beta=-1.83$ at $z=4$ to $-2.37$ at $z=7$, and that fainter galaxies have steeper $\beta$, while \citet{dun12} concluded a constant $\beta\sim-2.0$ at $z=5-7$, irrespective of redshift and UV magnitude.
\citet{ouc08} concluded that $\beta$ of LAEs, which are generally less dusty and/or younger, is slightly smaller than those of LBGs.
% popIIIのbeta?
When we changed the assumption $\beta=-2.0$ to $\beta=-2.5$ ($-1.5$), we found that $f^{phot}_{cont}$($0.92\micron$) varies as $7.01\pm3.44$ ($7.24\pm3.56$) $\times10^{-21}$ergs s$^{-1}$ cm$^{-2}$ \AA$^{-1}$. 
The possible scatter of $f^{phot}_{cont}$($0.92\micron$) due to the $\beta$ assumption is smaller than its internal error.
We also found that $f^{phot}$(Ly$\alpha$) remains unchanged for various assumed values of beta.

\begin{deluxetable*}{rrrrrrrrrrrr}
\tabletypesize{\scriptsize}
%\rotate
\tablecaption{Photometry of SDF-LEW-1 \label{tab_phot}}
\tablewidth{0pt}
\tablehead{
%\colhead{Coordinate\tablenotemark{a}} & 
\colhead{$B$} & \colhead{$V$} &  \colhead{$R$} & \colhead{$i'$} & \colhead{$NB921$} & \colhead{$z'$} & \colhead{$J$} & \colhead{$K$} & \colhead{$3.6\mu$m} & \colhead{$4.5\mu$m} & \colhead{$5.8\mu$m} & \colhead{$8.0\mu$m} 
}
\startdata
%13:24:57.98 27:23:49.4 & 
$>28.58$ & $>27.85$ & $>28.35$ & $>27.72$ & $24.69\pm0.024$ & $26.93\pm0.082$ & $>24.74$ & $>24.13$ & $>25.67$ & $>25.90$ & $>23.98$ & $>23.64$ 
% see toshikawa 2/15 mail
\enddata
\tablecomments{The lower limit denotes the $3\sigma$ limiting magnitude. The (limiting) magnitudes in optical and near-infrared bands are measured in $2$\arcsec aperture, while those in mid infrared bands are measured in $2\times$FWHM aperture. Refer to the photometric data descriptions to \citet{kas04} ($B$, $V$, $R$, $NB921$), \citet{poz07, gra11}($i'$, $z'$), Hasashi, M. et al. in prep., \citet{tos12}($J$, $K$), and Egami, E. et al. in prep. ($3.6\mu$m-$8.0\mu$m).}
%\tablenotetext{a}{J2000.0 equinox.}
\end{deluxetable*}

% \clearpage

It is particularly worth noting that the object has an extremely large EW$_0$ of $436^{+422}_{-149}$\AA, which is exceptionally larger than those of other LAEs at $z=6.5$ \citep{kas11} and at lower-$z$ \citep{shi06, gro07, ouc08}.
For example, \citet{gro07} showed that the EW$_0$ distribution of LAE at $z=3.1$ follows an exponential curve with an e-folding scale length of $76^{+11}_{-8}$\AA.
Several Pop III candidates with Ly$\alpha$ EW$_0$ as large as $\sim100$--$300$\AA~have been identified in previous works (e.g., \citealp{mal02}), though most of their EW$_0$ were only given by a lower limit because their continuum emissions were not detected in broad-band images.
\citet{ada11} found three LAEs with EW$_0>240$\AA~among their wide-field HETDEX (Hobby-Eberly Telescope Dark Energy Experiment) pilot spectroscopic survey, though one had no counterpart in the broad-band image, and the other two seemed to be extended Ly$\alpha$ blobs.
High EW$_0$ objects with no counterparts in the broad-band cannot be distinguished from noise contaminations \citep{ada11} or transient objects \citep{shi12}.
However, SDF-LEW-1 was actually detected in the $z'$-band image by virtue of deep SDF imaging data, and its EW$_0$ was reliably determined.
The 2nd largest EW$_0$ of our $z'$-detected LAE sample at $z=6.5$ was EW$_0=152$\AA~.
It should be noted that the $z'$-band image, which was constructed from $30$ hrs integration time in total by stacking all the data taken between 2001 and 2008, had a $3\sigma$ limiting magnitude as deep as $27.09$ mag.

\subsection{Estimate of intrinsic EW$_0$}

It is well known that Ly$\alpha$ photons are easily absorbed by dust and H {\sc i} clouds inside a galaxy.
In our previous study \citep{kas06}, we showed that the blue-side line profile of the Ly$\alpha$ emission of the composite spectrum of our LAE sample at $z=6.5$ can be simply explained by spectral broadening alone, which means that the blue-side of the observed Ly$\alpha$ emission is almost completely absorbed.
We confirmed that the blue-side of Ly$\alpha$ line profile of SDF-LEW-1 was also steep enough to be explained by simple spectral broadening (the instrumental resolution FWHM$=2.5$\AA~), though the spectral data quality of a single object was not good enough to accurately estimate the profile.
This is expected from the IGM attenuation model by \citet{mad95}, which predicts only $2\%$ transmission on the blue half of the Ly$\alpha$ emission line at $z=6.5$.
In this case, SDF-LEW-1 has an intrinsic EW$_0^{int}$ of $872^{+844}_{-298}$\AA.
The IGM/ISM absorption correction that we applied here is relatively large, and requires an enough attention on its uncertainty.
It has been suggested that large-scale outflows are ubiquitous in LAEs with Ly$\alpha$ velocity offsets of $\Delta v_{Ly\alpha}\sim100-300$ km/s \citep{mcl11, fin11a, has12} from the systemic redshift of the galaxy.
The outflow forces Ly$\alpha$ photons to reach frequencies further from line center to escape; therefore in this case, smaller IGM correction is expected.
%The redshifted Ly$\alpha$ photons can easily escape against the foreground gas attenuation; therefore in this case, smaller IGM correction is expected.
For example, when assuming an intrinsically symmetric line profile with $\Delta v_{Ly\alpha}=100$km/s, only $\times1.38$ correction instead of $\times2$ should be applied for SDF-LEW-1.
However, as discussed above, SDF-LEW-1 has a sharp blue edge of Ly$\alpha$ line profile, which is difficult to be reconciled with the strong outflow hypothesis \citep{ors12, lau11}.
On the other hand, the entire Ly$\alpha$ line at redshifts during the reionization epoch may be suppressed by a damping wing of the neutral IGM along the line-of-sight \citep{hai05}.
Therefore, the estimate of EW$_0^{int}$ might give only a lower-limit, if it was taken during the reionization epoch.
For simplicity, we here assume that the IGM is almost completely ionized (the neutral fraction of IGM hydrogen, $x_{\rm H I}\sim0$) at $z\sim6.5$.
In addition, there are several uncertainties in deriving EW$_0^{int}$ due to a possible variation of IGM attenuation from object to objet and internal ISM attenuation.
Although we hereafter use simple $\times2$ correction, it should be noted that EW$_0^{int}$ could have uncertainties, that are difficult to quantify, other than its observational errors.

The Ly$\alpha$ EW$_0^{int}\sim900$\AA~of SDF-LEW-1 is far greater than those of almost all LAEs ever observed.
Such an enormous EW$_0^{int}\sim900$\AA~in the Ly$\alpha$ emission cannot be attained by Population II synthesis.
\citet{dij07} also suggested that a Pop III contribution is inevitably required to self-consistently reproduce large EW$_0$ and observed Ly$\alpha$/UV luminosity functions at $z=6.5$ \citep{kas06}.
Although Pop III formation might continue down to $z=2.5$, depending on feedback efficiency \citep{tor07}, the expected fraction of Pop III objects among LAE sample significantly increases with redshift \citep{sca03}.
The probability to find Pop III objects is expected to be $100$ times larger at $z=6.5$ than at $z\sim3$.
SDF-LEW-1, which has an extraordinary EW$_0^{int}\sim900$\AA~at the high-redshift of $z=6.5$, can easily convince us that it is a plausible Pop III candidate.
The most promising way to identify this candidate as a Pop III dominant object is to directly detect its strong He {\sc ii} $\lambda1640$ emission.
The finding of such a rare object with a large EW$_0$ is really indebted to the Subaru wide-field data.
The target is the only plausible and accessible Pop III candidate among the largest sample of spectroscopically identified LAEs at $z=6.5$, which is very close to their birth epoch; {\it i.e.}, the end of the ^^ ^^ dark age".
%The SDF-LEW-1 is currently the most convincing target for Pop III because its Ly$\alpha$ EW$_0$ was surely mesaured.
The He {\sc ii} lines suffer minimal effects of scattering by gas and decreasing attenuation by intervening dust.
Moreover, the He {\sc ii} emission is expected to be detected at $12362$\AA, which fortunately does not correspond to the wavelengths of any strong OH sky lines.

\section{NIR spectroscopy}

We took deep NIR spectroscopy for SDF-LEW-1 with Subaru/MOIRCS \citep{ich06} on April 15-16 2011 (UT).
%to identify its redshifted He {\sc ii} $\lambda1640$ emission line at $\sim12374$\AA~
The observations were made with the $zJ500$ grism with a $0.8$ arcsec slit width ($R\sim450$).
The spectra covered $0.9-1.8\mu$m, with a pixel resolution of $5.57$ \AA.
We used MOS mode, which allows a more secure method of quickly and accurately aligning the slit on such a faint target, compared with the longslit mode in the case of MOIRCS.
We used two masks, and both masks contained the target.
%The spatial resolution was $0\arcsec.3$ pixel$^{-1}$ with $3$-pixel on-chip binning.
The total integration time was 44.8 ks ($\sim12.44$ hr), with a single exposure time of $500-900$ sec, depending on the sky background level.
During the observation, the telescope was nodded at two positions (A and B), with a dithering of $5.\arcsec0$ to achieve adequate background subtraction.
The seeing size was $\sim1$ arcsec.
We obtained a spectrum of the spectrophotometric standard star Feige 34 for flux calibration.
The data were reduced in a standard way using IRAF.
The spatial position of the blind target of each exposure spectrum was carefully determined by dithering the pattern of other bright objects on other slits in the same mask.
The final spectrum was constructed from the median frame with weights based on the relative flux transparency of each exposure, which was also measured from other bright objects in other slits.
% (hayashi et al.2011)First, bad pixels and cosmic rays were removed from each frame, and a A.B frame was created from a pair of successive frames observed at the two positions A and B.
%Then, flat-fielding was done with a dome-flat image for the individual A.B images. 
% Next, distortion was corrected using a calibration data provided by the MOIRCS instrument team. 
% After extracting a spectrum at each slit, wavelength calibration was done with the OH airglow lines. 
% Then, residual sky subtraction was carried out, since the A.B procedure alone might not completely remove the sky background due to its time variation. 
% All the spectra for each [OII] emitter were then co-added and an one-dimensional spectrum was extracted by combining ten pixels along a slit. 
% Finally, the telluric absorption and the instrumental efficiency were corrected using the spectra of BD+17.4708. As an error, the sky noise was estimated as a square root of the photon count on the sky spectrum.

\section{Results}

\begin{figure}
\epsscale{1.25}
\plotone{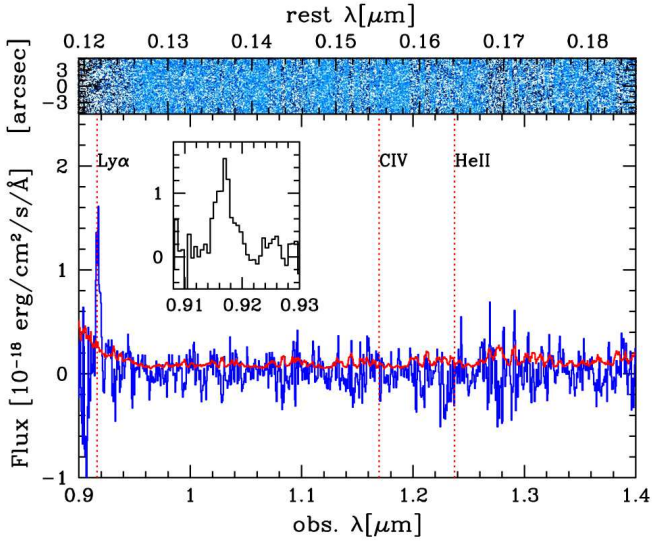}
\caption{Two-dimensional (upper) and one-dimensional (lower) NIR spectrum of SDF-LEW-1. 
The vertical dotted lines indicate the wavelength of Ly$\alpha$, C {\sc iv}, and He {\sc ii} emission lines at $z=6.538$.
The $1\sigma$ error spectrum (red line) is overlaid.
The inset panel shows the closeup of detected Ly$\alpha$ emission line. 
\label{fig_nirspec}}
\end{figure}

The final spectrum is shown in Figure \ref{fig_nirspec}.
The prominent Ly$\alpha$ emission was detected at $9169.6$\AA, which is almost consistent with the optical spectrum shown in Figure \ref{fig_optspec}, even at the bluest edge, where MOIRCS sensitivity drops significantly.
$S_W$ was estimated to be $S_W=8.26\pm1.16$, demonstrating an apparent asymmetric Ly$\alpha$ line profile.
The Ly$\alpha$ emission flux was estimated to be $f^{spec}_{NIR}$(Ly$\alpha$)$=(4.8\pm0.20)\times10^{-17}$ergs s$^{-1}$ cm$^{-2}$, which is higher than that measured in the optical spectrum.
The absolute flux calibration of the NIR spectrum is unreliable at the bluest wavelength; therefore, we hereafter refer the spectroscopic Ly$\alpha$ emission flux to that measured in the optical spectrum.

\begin{figure}
\epsscale{1.20}
\plotone{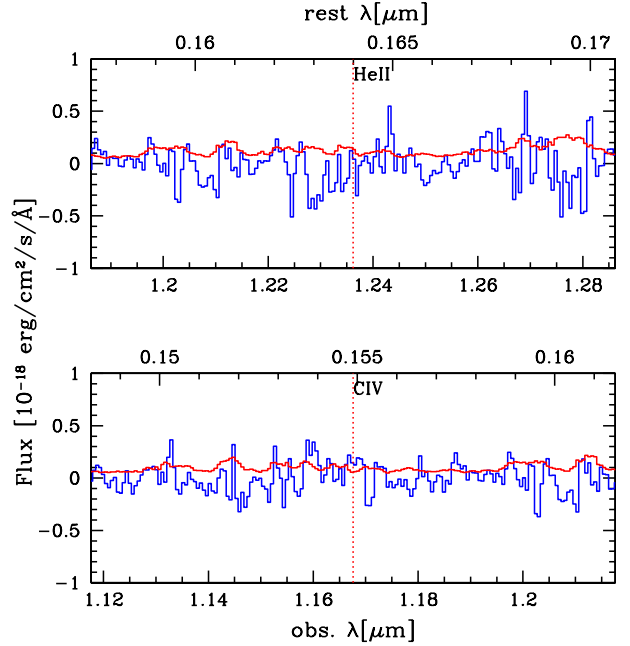}
\caption{The closeup spectra around expected wavelength of He {\sc ii} (upper) and C {\sc iv} (lower) at at $z=6.538$, respectively. 
The $1\sigma$ error spectrum (red line) is overlaid.  
\label{fig_he2c4}}
\end{figure}

No emission line feature is seen around $1.2362$\micron, at which the He {\sc ii} $\lambda1640$ emission is expected to appear, both on the one-dimensional (upper panel of Figure \ref{fig_he2c4}) and two-dimensional spectrum.
Note that the expected wavelength of He {\sc ii} appearance is based on the redshift determined from the peak of the Ly$\alpha$ emission, which could be, however, offset to the red from the systemic redshift due to absorption, though the upper limits described below would not significantly change even in the case.
The He {\sc ii} flux limits were derived using the $1.\arcsec0\times25.6$\AA~aperture, assuming that the He {\sc ii} line is expanded to almost seeing size and is unresolved.
The $3\sigma$ upper limit on the He {\sc ii} flux from SDF-LEW-1 is $3.39\times10^{-19}$ ergs s$^{-1}$ cm$^{-2}$, which corresponds to a luminosity of $1.66\times10^{41}$ ergs s$^{-1}$.
The UV continuum emission flux density at $1.2362\micron$ is photometrically estimated to be $f^{phot}_{cont}$($1.24\micron$)$=(3.91\pm2.07)\times10^{-21}$ ergs s$^{-1}$ cm$^{-2}$ \AA$^{-1}$ for $\beta=-2.0$, providing the $3\sigma$ upper limit on EW$_0$ of He {\sc ii} $\lambda1640$ as $\leq11.5^{+12.9}_{-3.98}$\AA.
The possible $\beta$ uncertainty could change $f^{phot}_{cont}$($1.24\micron$) to $3.32\pm1.64$ ($4.62\pm2.41$) $\times10^{-21}$ ergs s$^{-1}$ cm$^{-2}$ \AA$^{-1}$, for $\beta=-2.5$ ($-1.5$).
The uncertainty is still within the internal error of $f^{phot}_{cont}$($1.24\micron$) itself, though it is larger than in the case of $f^{phot}_{cont}$($0.92\micron$) due to a long extrapolation from the $z'$-band.
Equivalently, the $3\sigma$ upper limit on the observed He {\sc ii} $\lambda1640$/Ly$\alpha$ ratio is constrained to be $1.55\times10^{-2}$.
The ratio could become much smaller if the observed Ly$\alpha$ was attenuated by either the interstellar or intergalactic medium.
In the spectrum, the UV continuum flux was not detected with enough significance to measure its EW$_0$ spectroscopically.
The $3\sigma$ upper limit flux density at $0.95\micron$ is $6.24\times10^{-20}$ ergs s$^{-1}$ cm$^{-2}$ \AA$^{-1}$, which is higher than the photometric estimate.
It is generally very difficult to measure the EW$_0$ of high-z LAEs from their spectroscopic data, because most LAEs are too faint to accurately determine their continuum flux on the spectra.
%The $1\sigma$ lower limit of EW$_0$ was constrained to be EW$_0>xxx$, which is consistent to the photometric estimate of EW$_0=xxx$.

The C {\sc iv} $\lambda1549$ (lower panel of Figure \ref{fig_he2c4}) was neither detected around $1.1676$\micron, where the $3\sigma$ upper limit constraint is $2.83\times10^{-19}$ ergs s$^{-1}$ cm$^{-2}$ with the same aperture size as in the case of He {\sc ii}, suggesting that the object is not an apparent active AGN.
$f^{phot}_{cont}$($1.17\micron$) was estimated to be ($4.39\pm2.17)\times10^{-21}$ ergs s$^{-1}$ cm$^{-2}$ \AA$^{-1}$ for $\beta=-2.0$, and the $3\sigma$ upper limit on EW$_0$ of C {\sc iv} $\lambda1549$ is given by $\leq8.55^{+8.36}_{-2.82}$\AA.

\section{Discussion}

In this section, we discuss a variety of possible mechanisms that can explain the extremely strong Ly$\alpha$ emission with EW$_0^{int}=872^{+844}_{-298}$\AA~and no detection of He {\sc ii} $\lambda1640$ or C {\sc iv} $\lambda1549$ emissions from SDF-LEW-1.

\subsection{Pop III }

The large EW$_0^{int}=872^{+844}_{-298}$\AA~of the Ly$\alpha$ emission line of SDF-LEW-1 can be expected as an interesting characteristic of young, metal-poor or metal-free stellar populations in the first galaxies.
The derived upper limit on the He {\sc ii} $\lambda1640$ emission can constrain the IMF, metallicity, star formation history, and age of the population.
We here compare the observed EW$_0$ of the Ly$\alpha$ emission as well as the upper limit on the He {\sc ii} emission with the model predictions of \citet{rai10}, which are the extended calculations of \citet{sch03} and predict the EW$_0$ of these lines on the basis of their evolutionary synthesis code for a variety of different IMFs covering metallicities from zero to solar.

\begin{figure}
\epsscale{1.2}
\plotone{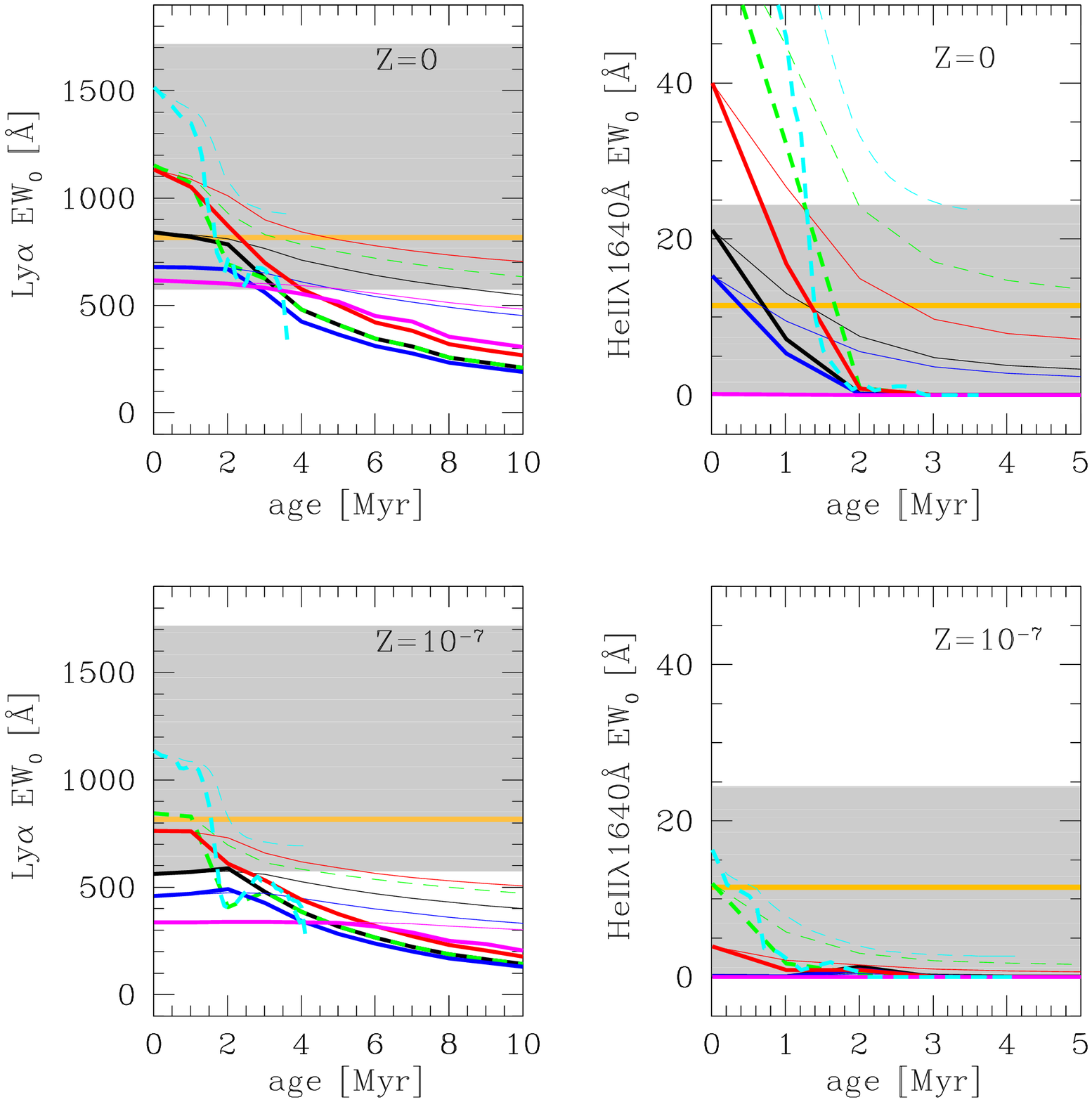}
\plotone{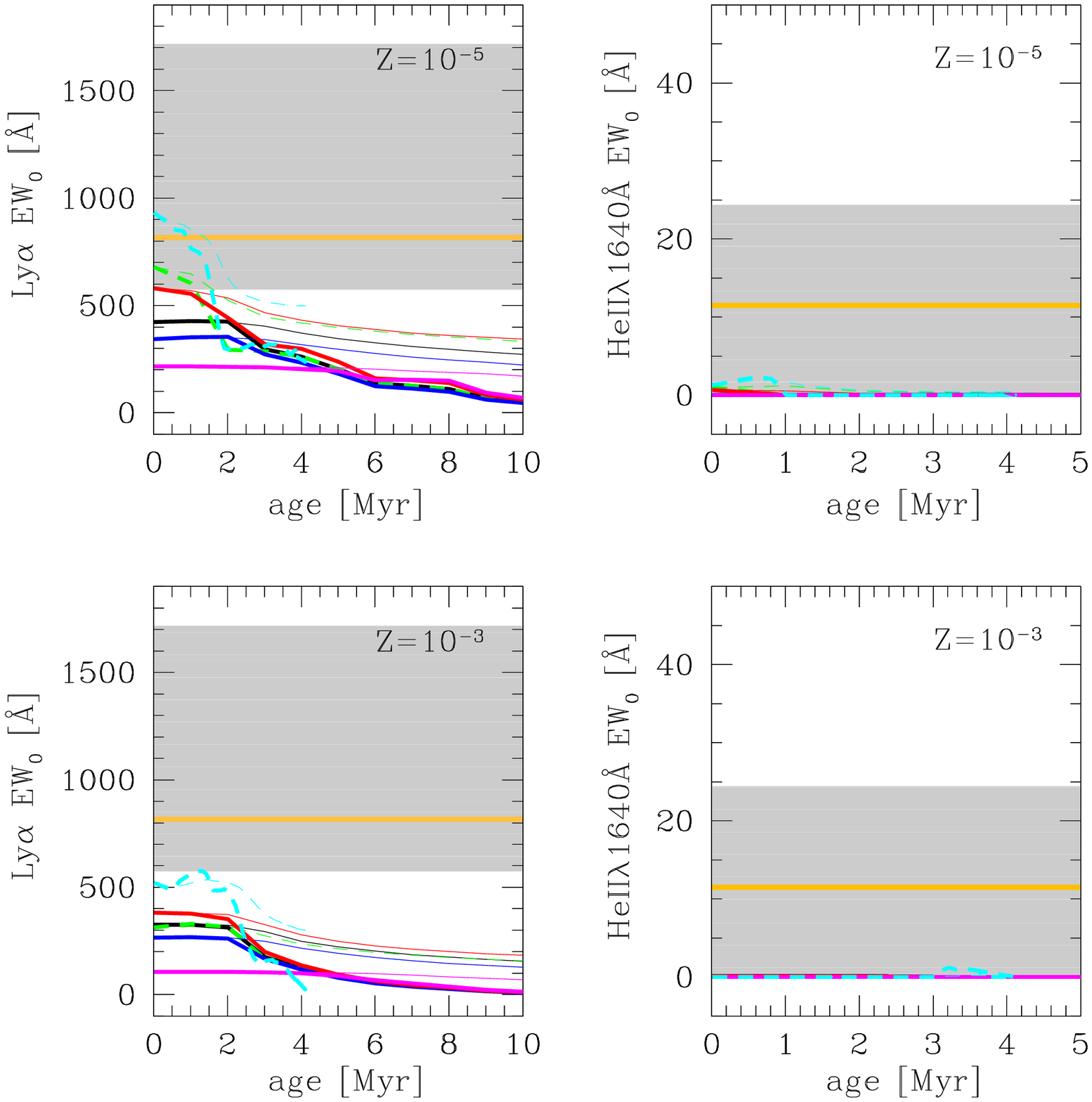}
\caption{Predicted Ly$\alpha$(left) and He {\sc ii} $\lambda1640$ (right) EW$_0$ as a function of age (Myr) for constant star formation (thin line) and young bursts (thick) models. 
The color codes of different IMF models are the same as those in Table 1 of \citet{rai10}: {\it i.e.}, 1) Model-S, Salpeter IMF with stellar mass range $1<M<100M_\odot$ (black); 2) Model-B, Salpeter IMF with $1<M<500M_\odot$ (green, dashed); 3) Model-C, Salpeter IMF with $50<M<500M_\odot$ (cyan, dashed); 4) Model-Sc, Scalo IMF with $50<M<500M_\odot$ (blue); 5) Model-TA, log-normal IMF with $1<M<500M_\odot$, M$_c=10$, and $\sigma=1.0$ as defined in \citet{tum06} (red); and 6) Model-TB, log-normal IMF with $1<M<500M_\odot$, M$_c=15$, and $\sigma=0.3$ (magenta).
For simplicity, we did not show the results from Model-L05 and Model-TE in \citet{rai10}, which are almost the same as those of the other models.
The Model-C, which contains only massive stars above $50$M$_\odot$, is calculated to the age $\simeq4$Myr.
The horizontal orange lines with shaded regions indicate the observational constraint of EW$_0$ (EW$_0^{int}=872^{+844}_{-298}$\AA~for Ly$\alpha$ and $\leq11.5^{+12.9}_{-3.98}$\AA~for He {\sc ii}) and their error ranges.
\label{fig_pop3model}}
\end{figure}

Figure \ref{fig_pop3model} compares the predicted EW$_0$ of Ly$\alpha$ and He {\sc ii} $\lambda1640$ as a function of age for models of constant star formation (CSFR) over 1 Gyr and young bursts ($<$4Myr).
We show six different IMF models: 1) Model-S, Salpeter IMF with stellar mass range $1<M<100M_\odot$ (black); 2) Model-B, Salpeter IMF with $1<M<500M_\odot$ (green, dashed); 3) Model-C, Salpeter IMF with $50<M<500M_\odot$ (cyan, dashed); 4) Model-Sc, Scalo IMF with $50<M<500M_\odot$ (blue); 5) Model-TA, log-normal IMF with $1<M<500M_\odot$, M$_c=10$, and $\sigma=1.0$ as defined in \citet{tum06} (red); and 6) Model-TB, log-normal IMF with $1<M<500M_\odot$, M$_c=15$, and $\sigma=0.3$ (magenta).
The color codes for the different IMF models are the same as those used in Table 1 of \citet{rai10}.
For simplicity, we did not show the results from their L05 and TE models, which are almost the same as those of the other models.
Note that the Model-C evolutionary track is only available up to $\sim4$Myr, because its IMF contains only massive stars.
The horizontal orange lines with shaded regions indicate the observed constraints of the EW$_0$ ($3\sigma$ upper limits for He {\sc ii}) and their error ranges.

Table \ref{tab_model} summarizes the acceptable age range in which each model can simultaneously explain the observed constraints of both Ly$\alpha$ and He {\sc ii} for each metallicity.
Figure \ref{fig_pop3model} exhibits the general trend that higher EW$_0$ are expected for both Ly$\alpha$ and He {\sc ii} lines at younger ages and lower metallicities.
The maximum EW$_0$ $\sim1500$\AA~and $\sim90$\AA~for Ly$\alpha$ and He {\sc ii} can be attained with $Z=0$, a very young age ($<$1Myr), and IMFs favoring more massive stars.
At zero-metallicity (the top panel of Figure \ref{fig_pop3model}), the large Ly$\alpha$ EW$_0^{int}=872^{+844}_{-298}$\AA~of SDF-LEW-1 can be explained by most of the IMF models at very young ages $<4$Myr, and by some of the CSFR models (Model-S, B and TA) even at ages $>4$Myr, while some models predict higher He {\sc ii} EW$_0$ than the observed upper limit at ages $<1$Myr.
All of the burst models have narrow age ranges, in which they are consistent with the constraints of both Ly$\alpha$ and He {\sc ii}.
The CSFR S model at ages $<9$Myr, the Sc model at ages $<5$Myr, and the TB model at ages $<5$Myr are also consistent with the observations.
At metallicity of $Z=10^{-7}$, some models can attain Ly$\alpha$ EW$_0$ as high as the observed, but only in a young population with ages $<$2-4Myr, while none of the predicted He {\sc ii} EW$_0$ are as high as in the case $Z=0$.
Model-B and Model-C are consistent with the observations at ages $<$1.5Myr and ages $<$4Myr for the burst and CSFR models, respectively.
Model-TA is consistent with the observations at ages $<$2.5Myr and ages $<$5.5Myr for the burst and CSFR models, respectively.
At metallicity $Z=10^{-5}$, the predicted He {\sc ii} EW$_0$ of all models are low enough to satisfy the observational constraint, and only Model-B and Model-C at very young ages at $<1-2$Myr are consistent with the observed Ly$\alpha$ EW$_0$.
Model-S, which has an upper mass as small as $100M_\odot$, is no longer consistent with the observations at $Z>10^{-5}$.
At metallicity $Z=10^{-3}$ and higher, no model can explain the observed large Ly$\alpha$ EW$_0$.

To summarize, some IMF models with $Z<10^{-5}$, and even $Z=0$, can simultaneously explain the large Ly$\alpha$ EW$_0$ and the stringent upper limit on the He {\sc ii} $\lambda1640$ EW$_0$, but only over a short ($\sim$ several Myr) age period.
Among the Salpeter IMF models (S, B, and C), models with higher upper mass ($500M_\odot$; Model-B and C) are able to explain the observation over a wider range of metallicities and ages.
% more preferable
The observed large EW$_0$ of Ly$\alpha$ is hardly explained by models with $Z\geq10^{-3}$.

\begin{deluxetable}{llrrrr}
\tabletypesize{\footnotesize}
%\rotate
\tablecaption{Constraints on Raiter et al.(2010)'s IMF models \label{tab_model}}
\tablewidth{0pt}
\tablehead{
\colhead{Z} & & \colhead{0} & \colhead{$10^{-7}$} & \colhead{$10^{-5}$} & \colhead{$10^{-3}$} 
}
\startdata
Model-S  & burst & $<3.5$Myr    & $<2$Myr   & -         & - \\
         & CSFR  & $<9$Myr      & $<2$Myr   & -         & - \\
\tableline
Model-B  & burst & $1.5-3.5$Myr & $<1.5$Myr & $<1$Myr   & - \\
         & CSFR  & $2$Myr$<$    & $<4.5$Myr & $<1.5$Myr   & - \\
\tableline
Model-C  & burst & $1.5-3.5$Myr & $<1.5$Myr & $<1.5$Myr & - \\
         & CSFR  & -            & $<4$Myr   & $<2$Myr   & - \\
\tableline
Model-Sc & burst & $<3$Myr      & -         & -         & - \\
         & CSFR  & $<5$Myr      & -         & -         & - \\
\tableline
Model-TA & burst & $1-4$Myr     & $<2.5$Myr   & -         & - \\
         & CSFR  & $1.5$Myr$<$  & $<5.5$Myr   & -         & - \\
\tableline
Model-TB & burst & $<3$Myr      & -         & -         & - \\
         & CSFR  & $<5$Myr      & -         & -         & - \\
%\enddata
\end{deluxetable}

% modelの不定性　caseB 最近ではtop-heavyにならないという説も。
%We note that the observed $L_{\rm Ly\alpha}$ did not correct for extinctions of both ISM and IGM, which could reduce more than half of the Ly$\alpha$ flux \citep{hai02}.
% Inoue: Note that our EWs are quantitatively very consistent with those with the same IMF and a similar duration in Raiter et al. (2010).
The models we used assume the conventional case B, which is not a good approximation for line and continuum emissions from a primordial nebula with low metallicity.
The Ly$\alpha$ line is boosted by $20-40\%$ when allowing for possible departures from case B, due to collisional effects, which increase the population of the n=2 level of hydrogen, leading to additional ionization, as discussed in \citet{rai10}.
Collisional excitation would be especially significant at low metallicity below $<0.03Z_\odot$ because of reduced radiative cooling and the consequent high electron temperature in the nebula.
The maximum Ly$\alpha$ EW$_0$ could reach $\sim2000$\AA~or higher at zero-metallicity when taking account of the departures from case B.
In contrast, the He {\sc ii} EW$_0$ could become weaker than in case B due to its dependence on the ionization parameter 
%(***Stasinska-Tylenda***) 
and enhanced nebular continuum emission.

In addition, the IMF of a metal-poor or metal-free gas is still subject to large theoretical uncertainties.
%Also, our knowledge of the IMF of metal-poor or free gas still remains very limited.
Based on simple physics, the primordial gases at the centers of dark matter minihalos of mass $\sim10^6M_\odot$ are condensed by molecular hydrogen cooling, becoming unstable to trigger a gravitational fragmentation with a very high typical mass of $\sim100M_\odot$, which is elucidated by many theoretical simulations \citep{abe02, bro02, yos06, osh06}.
The IMF models in \citet{rai10} also assume extremely top-heavy IMF; however, 
recent calculations (e.g., \citealp{kru09}, \citealp{tur09}, \citealp{sta12}) propound a more complex scenario with disk fragmentation under radiative feedback, resulting in a binary system or a small cluster of stars, with a final mass of only $\lesssim50M_\odot$.
\citet{hos11} found that Pop III stars grow up to only $\sim40M_\odot$ due to shut-off by radiative protostellar feedback.
In Figure \ref{fig_pop3model}, Model-S with smaller upper mass than Model-B and C has some difficulty in explaining the observed Ly$\alpha$ and He {\sc ii} lines at higher metallicities.
In the sense, the Ly$\alpha$ and He {\sc ii} lines of Pop III stars are expected to be smaller than those seen in \citet{rai10}, making it more difficult to consistently explain both observed line strengths at $Z>10^{-7}$.

%The current discussion also includes an uncertainty in the context of the Pop III-dominated fraction in this galaxy.
In this scenario, we can furthermore examine the stellar mass contributed by metal-free and poor stars using Ly$\alpha$ line luminosity.
%The observed EW$_0$s are consistent with most of zero or poor metallicity models with the age $<2-4$ Myr, which is much less than the dynamical timescale for galaxy-sized objects.
In \citet{rai10}, they predict the $1M_\odot$- and $1M_\odot$/yr- normalized Ly$\alpha$ fluxes, for burst and CSFR models, respectively.
We have estimated the stellar mass composed of metal-free and poor stars by scaling these predicted Ly$\alpha$ flux to the observed Ly$\alpha$ flux, assuming all the Ly$\alpha$ flux is due to these primitive stars.
The stellar mass was estimated to be the order of $10^6-10^7 M_\odot$ at the maximum age for acceptable models consistent with observed EW$_0$.
%It is significantly less massive than previously observed LAEs at high redshift.
This is larger than a characteristic stellar mass, $1.4\times10^5M_\odot$, of globular clusters \citep{har91}, and comparable to the mass of the most massive super star clusters \citep{kor09} or the least massive dwarf galaxies \citep{mat98} in the local universe.
% super star cluster 10^7 http://iopscience.iop.org/0004-637X/697/2/1180/fulltext/
Does this result suggest that this object is a building block at an early stage of galaxy formation?
In this case, the star-formation would have to propagate across the galaxy faster than the metal enrichment from the exploding Pop III stars, such that the current star-formation is still made from metal-free gas. 
Unfortunately, more quantitative estimate of the timescale of star-formation is difficult due to the lack of its physical size measurement.
Otherwise, our assumption above was incorrect, and part of the Ly$\alpha$ flux of this object could be due to normal stellar population.
\citet{ono10} have constructed the stacked SED of a large sample of LAEs at $z=5.7$ and $6.5$, and have determined the average stellar mass as small as ($3-10$)$\times10^7M_\odot$ with very young ages of $1-3$ Myr from the SED fitting including nebular emission.
\citet{pir07} have derived low stellar masses ($10^6-10^8M_\odot$) with very young (a few Myr) age for three faint LAEs at $5.2<z<5.8$ in the Hubble Ultra Deep Field (HUDF).
The observational estimate of stellar mass produced by normal stellar population in high-z LAEs still allows wide mass range; therefore it is difficult to estimate how much fraction Pop III stars are dominated in this galaxy.
%Given that SDF-LEW-1 had also a comparable total stellar mass, this galaxy should not be dominated by metal-zero/poor stars.
%Although the current starburst is still made from metal-free gas, there could be underlying old stellar population. 
%In the case, SDF-LEW-1 is not dominated by metal-zero/poor star.
It should be noted again that the observed Ly$\alpha$ flux is affected by dust and H {\sc i} absorptions; therefore our estimate of the stellar mass just gives the lower limit. 
%The current approach

%We caution that our analytical calculations do not account for all conceivable effects that could stem the growth of supermassive stars.

\subsection{Clumpy interstellar medium}

A large Ly$\alpha$ EW$_0$ might be the result of scattering in a clumpy, dusty interstellar medium (ISM).
In a clumpy ISM, Ly$\alpha$ photons scatter off at the edges of cold H {\sc i} gas clumps, in which dust is embedded, and survive to escape from the galaxy, while continuum photons are attenuated, leading to a relative enhancement of the Ly$\alpha$ EW$_0$ \citep{neu91, han06}.
Observational evidence has been found among evolved LAEs with large Ly$\alpha$ EW$_0$, possibly explained in terms of this mechanism by stellar population analysis \citep{fin08} and also supported by observations to account for the observed Ly$\alpha$/H$\alpha$ and H$\alpha$/H$\beta$ line ratio (\citealp{ate09, sca09a, fin11b}, but see \citealp{cow11}).

It would be interesting to know whether models can reproduce the extremely large EW$_0$ of the Ly$\alpha$ emission presented in the study.
Some theoretical models have attempted to address the quantitative estimate of Ly$\alpha$ EW$_0$ enhancement due to clumpy ISM, though it is challenging.
\citet{kob10} concluded that a clumpy dust distribution is required for their model in order to reasonably reproduce the Ly$\alpha$ luminosity function (LF), UV LF, and EW distribution of LAEs from $z=3.1$ to $6.5$.
In the model of \citet{kob10}, the dust clumpiness parameter, denoted by $q_d$\footnote{This parameter was originally introduced by \citet{fin08}, but their definitions are slightly different from each other.} was introduced and was defined as the optical depth ratio of Ly$\alpha$ to the continuum, which effectively reflects the interstellar dust geometry.
The best-fit value of this parameter was found to be $q_d=0.149\pm0.03$, implying a rather clumpy ISM distribution.
Assuming that $q_d=0.15$, the Ly$\alpha$ EW$_0$ enhancement factor, $\Gamma=$EW$_{Ly\alpha}^{obs}$/EW$_{Ly\alpha}^{int}$ attains $1.53$ at maximum, which is not large enough to practically account for the observed EW$_0\sim900$\AA~of SDF-LEW-1 with a normal stellar population even with clumpy ISM.
However, galactic-scale outflow, if present, could significantly reduce the scattering optical depth of Ly$\alpha$ in a low-density outflowing ISM.
In the case, $\Gamma$ could become as large as $\sim30$ at maximum when the dust content amounts to $A_V\sim4.5$ mag, which is , however, not likely for LAEs at $z\sim6.5$.
%In spite of its reality with several uncertain assumptions made in the modeling, it is difficult to completely exclude the possibility that the observed large Ly$\alpha$ EW$_0$ of SDF-LEW-1 is enhanced by clumpy ISM.
\citet{day11} also require a clumpy ISM to fit their model to the observed Ly$\alpha$ and UV LFs, enhancing the Ly$\alpha$ EW$_0$ by a factor of $1.3$, and $3.7$ when taking into account inflow/outflow.
%Within the uncertainties of models, 
Although several uncertain assumptions are made in the modeling, we cannot conclusively rule out the model that SDF-LEW-1 is composed of a normal stellar population with clumpy ISM.
Detailed radiative transfer simulations of the Ly$\alpha$ photons will be required to obtain a more quantitative estimate of Ly$\alpha$ EW$_0$ enhancement in a clumpy ISM.

%しかし、EWの大きなLAEを作るphase=starburst時には、f0->f0windに代わるので、AVが大きくなってもΓは大きくなり続ける。この場合LyAは無限に大きくなりうるが、escape fractionがdust吸収に無関係という仮定を置いているので、この仮定の妥当性はわからない。
%neverthless, この場合でも観測されうるLyAEWを持つ銀河は全体のｘｘ％である。

%The possible existence of this and other kinds of geometries has been reported to be able to recover the observed Ly/H and H/H ratios and the observed SEDs of LAEs at different redshifts (Finkelstein et al. 2011; Scarlata et al. 2009; Guaita et al. 2011; Finkelstein et al. 2008).

\subsection{AGN}

An alternative origin for a large Ly$\alpha$ EW is a large AGN contribution, but we did not identify a C {\sc iv} emission line as an apparent AGN signature.
Typical values of the line ratio for radio galaxies is Ly$\alpha$/C {\sc iv}$=6.7$ at $2<z<3$ \citep{hum08}, and Ly$\alpha$/C {\sc iv}$=8.6$ at $0<z<3$ \citep{mcc93}.
The ratio is expected to be smaller to be Ly$\alpha$/C {\sc iv} $\sim4$ in the case of narrow-line AGNs \citep{sch03}, and $\sim4.76$ for local Syfert galaxies \citep{fer86}.
These values are far inconsistent with the observed $3\sigma$ lower limit of Ly$\alpha$/C {\sc iv}$\geq82.7$.
In addition, the optical spectrum of SDF-LEW-1 does not reveal any signature of NV $\lambda1240$ emission line, which is a strong high-ionization metal line indicating AGN activity.
Although an AGN without accompanying strong metal-line emissions has been discovered \citep{hal04}, its Ly$\alpha$ emission is moderately broad and EW$_0=34$\AA, that are dissimilar to SDF-LEW-1.
We have also checked for possible time variability, which is one of the effective ways to identify AGN suggested by \citet{shi12}.
Based on the original SDF image \citep{kas04} with $8.4$ hr of integration time, taken during the 2002-2003 period, the $z'$-band magnitude was measured as $27.05\pm0.127$.
Within the margin of error, this is almost identical to that used in the study (Table \ref{tab_phot}).
Thus, we did not detect any apparent signatures of its variability.
We conclude that SDF-LEW-1 is most likely not photoionized by an active nucleus.

% AGNの典型的なHeII/LyA比？

\subsection{Ly$\alpha$ blob}

A large Ly$\alpha$ EW$_0$ can be naturally expected in the case of a spatially extended Ly$\alpha$ source, often referred to as ^^ ^^ Ly$\alpha$ blob" \citep{ste00, mat04}; however, the NB image of SDF-LEW-1 appears to be almost unresolved with FWHM$=1.\arcsec04$ (the PSF FWHM of $NB921$-image is $0.\arcsec98$), showing no evidence of a Ly$\alpha$ blob.
\citet{ouc09} discovered a Ly$\alpha$ blob at $z=6.6$.
Employing the same detection surface brightness threshold of $26.8$ mag arcsec$^{-2}$ used in their work, we found that the isophotal area, which is defined as the image area above the detection threshold at the object's position, of SDF-LEW-1 was $2.77$ arcsec$^2$, which is much smaller than the $5.22$ arcsec$^2$ obtained by \citet{ouc09}.
Also, when we used the fainter detection threshold of $28.0$ mag arcsec$^{-2}$ as \citet{mat04} in their systematic search for Ly$\alpha$ blobs, the isophotal area of SDF-LEW-1 was $6.32$ arcsec$^2$ ($\sim14$kpc in diameter at $z=6.5$), which is much smaller than $16$ arcsec$^2$ ($\sim30$kpc in diameter at $z=3.1$), critical criterion for a blob in \citet{mat04}.
We can conclude that SDF-LEW-1 is not a Ly$\alpha$ blob.

It is interesting to note that a Ly$\alpha$ blob sometimes \citep{pre09, sca09b} has a prominent He {\sc ii} emission, caused by either cooling gas accreting on the dark matter halo of the galaxy \citep{hai00, yan06, lat11}, or photoionization by a hard ionizing source, such as an AGN \citep{yan09} or very low-metallicity stars.

% Dijkstra 2009: compact cooling cloud

\subsection{Fluorescently illuminated Ly$\alpha$ emission}

\citet{can12} recently reported interesting evidence of Ly$\alpha$ emissions originating in fluorescently illuminated by strong radiation from the nearby hyperluminous quasar HE0109-3518 at $z=2.4$.
The sample contains several LAEs with large Ly$\alpha$ EW$_0>$240\AA~, which could be boosted by fluorescent emission, irrespective of internal star formation.
This could be another effective mechanism for enhancing the Ly$\alpha$ EW$_0$.
However, in contrast to their observation, there is no evidence of a systematic excess of LAEs with high EW$_0$, steepening Ly$\alpha$ LF, or spatially strong LAE clustering around SDF-LEW-1.
Two thirds of their sample have no stellar continuum counterparts, whereas SDF-LEW-1 was actually detected in the $z'$-band image, though most of the $z'$-band flux would have been contributed by Ly$\alpha$.
% little contribution of associated star-formation
Depending on its internal velocity field, the emission line profile of a fluorescent Ly$\alpha$ is ideally expected to be double-peaked \citep{can07, ade06}, which was not identified in the spectrum of SDF-LEW-1.
% 一番近いLAE
Finally, we did not identify apparent strong ionizing source candidates around SDF-LEW-1.
The closest LAE at $z=6.5$ is $6.1$ comoving Mpc along a projection away from SDF-LEW-1 and its $f^{phot}_{cont}$($0.92\micron$) is smaller than that of SDF-LEW-1.
The next closest is $8.5$ Mpc away, and its $f^{phot}_{cont}$($0.92\micron$) is only twice as large.
There are three $i'$-dropout objects \citep{tos12} within a separation of $10$ Mpc on a projection from SDF-LEW-1, all with a rest-UV magnitude of more than $-21$mag.
Hence, they do not seem to be strong radiation sources, if they were at $z=6.5$.
Therefore, it is very unlikely that SDF-LEW-1 is fluorescently illuminated by a nearby quasar.

\section{Summary and Implications}

We found an interesting LAE at $z=6.538$ with an extraordinary large Ly$\alpha$ EW$_0^{int}=872^{+844}_{-298}$\AA, which is exceptionally larger than those of other LAEs.
%This is an interesting candidate for very metal-poor and PopIII galaxies.
%The emission line property is generally determined by the age of the stellar population, the stellar initial mass function, the metal and dust content.
The continuum emission of the object was actually detected in the $z'$-band image, and its EW$_0$ was reliably determined.
Follow-up deep NIR spectroscopy detected neither He {\sc ii} $\lambda1640$ nor C {\sc iv} $\lambda1549$ emission lines from the object.
No detection of C {\sc iv} convinces us that it is unlikely that the object is being photoionized by an active nucleus. 
It has no apparent Ly$\alpha$ blob morphology features.
We obtained no positive evidence that supports the fluorescent boosting of the Ly$\alpha$ emission.
The observed extremely large EW$_0$ of the Ly$\alpha$ emission and upper limit on EW$_0$ of the He {\sc ii} $\lambda1640$ emission can be explained by population synthesis models favoring very young and massive metal-poor ($Z<10^{-5}$) stars, or even Population III stars.
The observed large EW$_0$ of Ly$\alpha$ is hardly explained by models with higher metallicities of $Z\geq10^{-3}$.
A large Ly$\alpha$ EW$_0$ with no He {\sc ii} emission can also be explained by clumpy ISM, which could enhance the Ly$\alpha$ EW$_0$.
We cannot draw a firm conclusion about the origin of the extremely large Ly$\alpha$ EW$_0$ of this object.
Dedicated follow-up observations, such as deep NIR/MIR imaging to determine its age or dust content, will be required to put further constraints on its origin.
It should be noted that the predicted UV continuum flux of SDF-LEW-1 is $\sim28.2$ mag., which is likely detectable by the Hubble Space Telescope (HST) and its observation would give more accurate estime of EW$_0$.

A combination of strong Ly$\alpha$ emission and He {\sc ii} $\lambda1640$ emission is the most promising indicator expected for Pop III-dominated galaxies.
However, the Ly$\alpha$ emission is a resonant line with a large cross section; therefore, its radiative transfer, which is complicated by the geometry and velocity field of the ISM, IGM, and dust attenuation, can significantly alter the observed Ly$\alpha$ emission.
%Appropriate modeling of the observed Ly$\alpha$ emission requires a complete understanding the radiative transfer of the Ly\alpha photones with simple assumptions of the dynamics and chemical composition of the ISM.
%It is interesting or challenging whether such a model can reproduce the large EW of the Ly$\alpha$ emission as presented in the study.
%reionizationになるとLyAが食われるので、line比とモデルとの比較はよく難しくなる
Pop III-dominated galaxies are more likely to be found at much higher redshifts, where the observed Ly$\alpha$ flux can be significantly reduced by neutral IGM during the reionization epoch.
The effect makes it complicated to apply such a diagnostic to find first galaxies.
Another possible diagnostic for the Pop III-dominated galaxies is the line ratio of He {\sc ii} $\lambda1640$ and H$\alpha$, which could be a good indicator of the IMF \citep{joh09}.

The He {\sc ii} emission line should be a unique signature of metal-poor and Pop III stars; however, for a single burst of star formation, the He {\sc ii} line is so short-lived ($<2$Myr) that its practical detection relies on surveying large volumes.
No detection of a distinct He {\sc ii} emission is not necessarily inconsistent with a Pop III interpretation because of its very short duration.
%predicted EW$_0$ of He {\sc ii} is practically detectable only from very young ($<$2Myr) and metal-poor ($Z<10^{-7}$) population.
Nonetheless, the He {\sc ii} $\lambda1640$ signature is likely one of the most promising indicators of metal-free stars and will continue to be sought using future large telescopes.
Detection of the He {\sc ii} line emitted from the first galaxies at $z\geq8$ will be made possible (e.g., \citealp{zac11}) in the next decade, with the James Webb Space Telescope (JWST) or Extremely Large Telescopes (ELTs).

\acknowledgments

We thank the referee, Steven Finkelstein, for his helpful comments that improved the manuscript.
We thank Daniel Schaerer, Mark Kobayashi, Mark Dijkstra, and Masashi Chiba for their useful discussions.
We are grateful to the Subaru Observatory staffs for their help with the observations.
We especially thank Ichi Tanaka with help on MOIRCS observing and its data reduction.
The observing time for part of this project was committed to all the Subaru Telescope builders.
This research was supported by the Japan Society for the Promotion of Science through Grant-in-Aid for Scientific Research 23340050.


%% in parentheses, after the keyword, but they will not be verified.

{\it Facilities:} \facility{Subaru (MOIRCS, FOCAS, Suprime-Cam)}, \facility{KeckII (DEIMOS)}, \facility{UKIRT (WFCAM)}, \facility{Spitzer (IRAC, MIPS)}.




\clearpage

%% the file, uncomment the \plotone, \plottwo, and
%% If you need a layout that cannot be achieved with \plotone or
%% \plottwo, you can invoke the graphicx package directly with the
%% \includegraphics command or use \plotfiddle. For more information,
%% please see the tutorial on "Using Electronic Art with AASTeX" in the
%% documentation section at the AASTeX Web site,
%% http://www.journals.uchicago.edu/AAS/AASTeX.
%%
%% The examples below also include sample markup for submission of
%% supplemental electronic materials. As always, be sure to check
%% the instructions to authors for the journal you are submitting to
%% for specific submissions guidelines as they vary from

%% This example uses \plotone to include an EPS file scaled to
%% 80% of its natural size with \epsscale. Its caption
%% has been written to indicate that additional figure parts will be
%% available in the electronic journal.

\clearpage

\epsscale{.80}
%\plotone{f1.eps}
%\caption{ The distribution of intrinsic rest-frame equivalent width (EW$_0^{int}$) of high-$z$ LAE sample at $z=6.5$ (red).
%The EW$_0^{int}$ distributions of other lower-$z$ LAE sample at $z=5.7$ (blue) in the SDF (Shimasaku et al. 2006), $z=5.7$ (magenta), $z=3.7$ (green), and $z=3.1$ (cyan) in the SXDS (Ouchi et al. 2008, normalized to arbitrary) are also shown for comparison.
%All the objects plotted here are detected in both broad- and narrow- band images, which gives a secure estimate for EW$_0^{int}$.
%Self-absorption due to H {\sc i} and dust inside a object are not corrected.
%One object at $820$\AA~bin (red shaded point) corresponds to SDF-LEW-1.\label{fig1}}
%\end{figure}

%\clearpage

\clearpage

\clearpage

\clearpage

\clearpage

\end{document}